\documentclass[12pt]{article}
\usepackage{a4wide}
\usepackage{graphics}
\usepackage{amsfonts}
\usepackage{amstext}
\usepackage{amsmath}
\usepackage{cite}
\renewcommand{\theequation}{\arabic{section}.\arabic{equation}}
\newcommand{\Section}[1]{\setcounter{equation}{0}\section{#1}}
\def\Tr{\mbox{Tr}}
\def\exp{\mbox{exp}}
\def\ln{\mbox{ln}}
\def\Prob{\text{Prob}}

\begin{document} 

\title{
One-Dimensional Partially Asymmetric Simple Exclusion Process 
on a Ring with a Defect Particle}
\author{Tomohiro SASAMOTO 
{\footnote {\tt e-mail: sasamoto@monet.phys.s.u-tokyo.ac.jp}
\vspace{5mm}} \\
{\it Department of Physics, Graduate School of Science,}\\
{\it University of Tokyo,}\\
{\it Hongo 7-3-1, Bunkyo-ku, Tokyo 113-0033, Japan}}
\date{}

\maketitle

\begin{abstract}
The effect of a moving defect particle for 
the one-dimensional partially asymmetric simple 
exclusion process on a ring is considered.
The current of the ordinary particles,
the speed of the defect particle and the
density profile of the ordinary particles 
are calculated exactly.
The phase diagram for the correlation length 
is identified.
As a byproduct, the average and the variance of 
the particle density of the one-dimensional partially asymmetric 
simple exclusion process with open boundaries
are also computed.

\end{abstract}

\Section{Introduction}
\label{intro}
Recently, the one-dimensional
asymmetric simple exclusion process (ASEP) 
has attracted much attention in various fields of science
including mathematics, physics and biology
\cite{L,L2,Sp,SZ}.
There are several reasons for this. 
First of all, the ASEP has a lot
of applications to realistic systems. For instance,
the ASEP can be considered as a simplest model of traffic 
flow \cite{SW,ERS,LPK}. 
It is also used to study the dynamics of interface \cite{HZ}. 
There are other applications as well.
Second, the ASEP shows rich non-equilibrium behaviors 
such as the shock wave \cite{DJLS,DLS,DGLS}, 
boundary induced phase transition \cite{Krug},
and the unusual dynamical scaling \cite{GS,Kim}.
Since there is no established framework for far-from
equilibrium systems, the ASEP has played an important role 
in non-equilibrium statistical physics.
Third, some properties of the ASEP can be studied exactly
by the Bethe ansatz or the so-called matrix product ansatz.
Since the effects of fluctuations become strong in low
dimensions, mean-field analysis sometimes fails to give 
correct answers. Exact solutions give us an insight 
how to understand the problems which can not be solved exactly.

The ASEP is a one-dimensional lattice gas model.
Each particle tends to hop to the right 
with rate $p$ and to the left with rate $q$. 
In addition, they interact with one another through 
hard-core exclusion interaction.
In this article, we consider the stationary state of the ASEP 
on a ring with one defect particle.
We sometimes refer to the particles as ordinary particles
to distinguish them from the defect particle.
We assume that the defect particle tends to hop 
to the right with rate $\alpha$ and that it
can exchange its position with an ordinary particle on the left 
nearest neighboring site with rate $\beta$. 
By rescaling time, we can set $p=1$ without loss of 
generality. 
Then the above rules are represented as
\begin{equation}
\begin{matrix}
   & 10 \rightarrow 01  &\text{with rate} & 1 \\
   & 01 \rightarrow 10  &\text{with rate} & q \\
   & 20 \rightarrow 02  &\text{with rate} & \alpha \\
   & 12 \rightarrow 21  &\text{with rate} & \beta . 
\end{matrix}
\label{rates}
\end{equation}
Here $0$, $1$ and $2$ denote an empty site, an
ordinary particle and the defect particle respectively. 
Several exact results have already been obtained 
for the $q=0$ case \cite{Mallick,DE99}. 
This case is referred to as the ``totally asymmetric''
case in the following.
The current of ordinary particles, the
speed of the defect particle, the correlation 
length and the current fluctuation have been computed 
exactly for this special case.
Much less is known for the ``partially asymmetric'' case where $q$ is 
not necessarily zero although some exact results were obtained in
\cite{Jafa}.
In this article, the model is studied for the $0<q<1$ case.
We compute the current of the ordinary particles,
the speed of the defect particle and the density profile of the 
ordinary particles seen from the defect particle.
We exploit the connection of the present model to the 
partially ASEP with open boundaries, which was recently solved by 
using the theory of the $q$-orthogonal polynomials
\cite{PASEP,PASEP2}. 
We do not consider the $q\geq 1$ case, in which the current becomes 
zero in the thermodynamic limit.  
We remark that, for the partially ASEP with open boundaries, 
the $q=1$ case and the $q>1$ case were studied in 
\cite{SEP} and \cite{BECE} respectively.

The paper is organized as follows.
In the next section, the stationary state
of the process is constructed by 
the so-called matrix product ansatz.
The current of the ordinary particles
and the speed of the defect particle are
calculated in section \ref{current}.
In section \ref{density},
the average density profile is computed. 
The phase diagram for the correlation length is 
identified.
The concluding remarks are given in the last section. 
In Appendix A, we compute the average and the 
variance of the particle density  
for the partially ASEP with open boundaries. 
This is a byproduct of the computations in the main text.

\Section{Exact Stationary State in Matrix Product Form}
The stationary state 
of the process can be constructed by applying
the so-called matrix product ansatz.
This technique was first introduced in \cite{HN} 
to study directed animals. 
It was then applied to the ASEP in \cite{DEHP}.
Since then this method has been successfully generalized and applied
to many interesting phenomena. See for instance the references 
in \cite{PASEP}.
In the formalism of the matrix product ansatz,
the stationary state of the process (\ref{rates}) is constructed as follows.
Let us denote the number of 
particles as $N$ and the lattice length as $L+1$. 
To be specific we explain the construction mainly by using 
an example for the case $L=2,N=1$: the case where 
there are one ordinary particle, 
one defect particle and one empty site on the chain with three sites.
For this example,
the stationary state can be written as a linear combination
of the six configurations 
$012,021,102,120,201,210$ and is completely determined 
by specifying the corresponding six coefficients 
$P(012),P(021),P(102),P(120),P(201),P(210)$.
Each coefficient $P(\tau_1,\tau_2,\tau_3)$
is the probability that the system
is in a configuration $ \tau_1\tau_2\tau_2 $
with $\tau_j=0,1,2 \, (j=1,2,3)$
for the stationary state.

\vspace{5mm}
\noindent
{\it Step 1}: 
To each configuration  $0,1,2$ at one site, we associate a matrix 
$E, D$ and $A$ respectively. The space on which these matrices
act is not specified at this stage. 
\begin{equation}
\begin{array}{ccc}
  0  &\leftrightarrow &E \\
  1  &\leftrightarrow &D \\
  2  &\leftrightarrow &A 
\end{array}
\end{equation}

\noindent
{\it Step 2}: 
To each configuration
$\tau_1\tau_2\tau_3$ of the whole lattice
we associate a matrix product. 
For instance the configuration $ 120 $ is associated with 
the matrix product $DAE$. 

\vspace{5mm}
\noindent 
{\it Step 3}: 
We construct a state, of which each coefficient is given by the 
trace of the matrix product.
For the present example, the six coefficients
$P^{(u)}(012),P^{(u)}(021),P^{(u)}(102),\\
 P^{(u)}(120),P^{(u)}(201),P^{(u)}(210)$ 
are given by
$\Tr(EDA),\Tr(EAD),\Tr(DEA),\Tr(DAE),\\ \Tr(AED),\Tr(ADE)$
respectively.
Here we denote the coefficients as 
$P^{(u)}(\tau_1,\tau_2,\tau_3)$ because the corresponding 
state is unnormalized. 
The coefficients of the normalized state 
$P(\tau_1,\tau_2,\tau_3)$ is obtained 
by simply dividing the unnormalized ones by 
the normalization constant,
\begin{align}
\label{norm_ex}
Z_{L=3,N=1}
&=
\Tr(DEA)+\Tr(DAE)+\Tr(EDA)
\notag\\
&\quad
+\Tr(EAD)+\Tr(ADE)+\Tr(AED).
\end{align}
In the above prescription,
there is a problem about the existence of trace. 
We will see that we can take a special form 
of the matrix $A$ which ensures the existence
of the trace.

\vspace{5mm}
\noindent
{\it Step 4}:  
We can show that 
the constructed state is the stationary state of the process
if the matrices $D,E,A$ satisfy the conditions,
\begin{gather}
  DE - q ED   = \zeta (D+E) , \notag\\
  \beta DA   = \zeta A ,     \notag\\
  \alpha AE   = \zeta A .  
  \label{alg}
\end{gather}
Here $\zeta$ is an arbitrary number and is taken to be
$\zeta = 1-q$ hereafter.
In addition, we set  $\tilde{\alpha}=\alpha/(1-q),\,
\tilde{\beta}=\beta/(1-q)$. 
The demonstration of the fact that the above 
algebraic relations are sufficient conditions for the stationarity
of the process proceeds in almost the same
manner as for the ASEP \cite{DEHP} and therefore is omitted.
As we will see at the end of the section, we can take a 
matrix $A$ of the form $A=|V\rangle \langle W|$, where 
$|V\rangle$ and $\langle W|$ are some vectors.
Then we notice that the above algebraic relations (\ref{alg})
are exactly the same as those for the partially ASEP with
open boundaries \cite{DEHP,PASEP}.

\vspace{5mm}
\noindent
{\it Step 5}: 
Obviously, the construction of the stationary state above 
works for any choice of $L,N$.
Explicitly, the coefficients
$P(\tau_1,\tau_2,\ldots,\tau_L)$
of the normalized stationary state is given by 
\begin{equation}
  \label{mpa_ce}
  P(\tau_1,\tau_2,\ldots,\tau_L)
  =
  \frac{1}{Z_{L,N}}
  \langle W|
  \prod_{j=1}^{L}
  [ \tau_j  D 
    +
   (1-\tau_j)  E ]
  |V\rangle , 
\end{equation}
for the configuration $\tau_1\tau_2\ldots\tau_L$
which satisfies $\sum_{j=1}^L\tau_j=N$.
Here the normalization constant $Z_{L,N}$ is 
\begin{equation}
  \label{norm}
  Z_{L,N}
  =
  \sum_{\scriptsize
     \begin{array}{c}
       \tau_1,\tau_2,\ldots,\tau_L\\
       \sum_{j=1}^L \tau_j = N\\
     \end{array}}
  \langle W|
  \prod_{j=1}^{L}
  [ \tau_j  D 
    +
   (1-\tau_j)E]
  |V\rangle .
\end{equation}
The summation is over $\tau_j=0,1$ for $j=1,2,\ldots,L$
with the condition $\sum_{j=1}^L \tau_j = N$.
Hence we have constructed the stationary state 
in matrix product form.

\vspace{8mm}
In this article, we will compute the current of the 
ordinary particles, the speed of the defect particle 
and the density profile of the ordinary particles
seen from the defect particle.
These quantities are written 
in terms of the matrices as follows.

\noindent
The current of the ordinary particles is given by 
\begin{align}
  J 
  &=
  \Prob (\tau_j=1,\tau_{j+1}=0)
  -  
  q\,\Prob (\tau_j=0,\tau_{j+1}=1)
  +
  \beta \,\Prob (\tau_j=1,\tau_{j+1}=2)
  \notag\\
  &=
  (1-q)
  \left[
  \frac{N}{L+1}
  \frac{Z_{L-1,N}}{Z_{L,N}}
  -
  \frac{L-N}{L+1}
  \frac{Z_{L-1,N-1}}{Z_{L,N}}
  \right] .
  \label{def_J}
\end{align}

\noindent 
The speed of the defect particle has a similar expression,
\begin{align}
  v 
  &=
  \alpha\,\Prob (\tau_j=2,\tau_{j+1}=0)
  -  
  \beta\,\Prob (\tau_j=1,\tau_{j+1}=2) 
  \notag\\
  &=
  (1-q)
  \frac{Z_{L-1,N}-Z_{L-1,N-1}}{Z_{L,N}}.
  \label{def_speed}
\end{align}

\noindent
The average density of the ordinary particles at site $j$
seen from the defect particle reads
\begin{align}
  \langle n_j \rangle_{L,N}^{(u)} 
  &=
  \sum_{\scriptsize
     \begin{array}{c}
       \tau_1,\tau_2,\ldots,\hat{\tau}_j,\cdots,\tau_L\\
       \tau_1 +\cdots + \hat{\tau}_j+\cdots+ \tau_L = N-1\\
     \end{array}}
  \langle W|
  \prod_{k=1}^{j-1}
  [ \tau_k  D 
    +
   (1-\tau_k)E]
  D
  \prod_{k=j+1}^{L}
  [ \tau_k  D 
    +
   (1-\tau_k)E]
  |V\rangle ,
  \label{def_density}
\end{align}
where $\hat{\tau}_j$ indicates that it is omitted.
This is an unnormalized quantity.
The normalized density is given by
$  \langle n_j \rangle_{L,N} 
  =
  \langle n_j \rangle_{L,N}^{(u)} /Z_{L,N}$.  

Here we note that the process has 
a particle-hole symmetry which reduces our computations
greatly.
The process is invariant 
when the particles and holes are interchanged, 
the direction of particle hoppings are reversed, 
and $\alpha$ and $\beta$ are exchanged at the same time.
In other words, the process is invariant under the change,
\begin{align}
\notag
\text{particle}
&\leftrightarrow
\text{hole}
\\
\label{symmetry}
\alpha
&\leftrightarrow 
\beta
\\
\notag
\text{site number} \,\, j
&\leftrightarrow 
\text{site number} \,\, L-j+1.
\end{align}
Due to this symmetry, it is sufficient to obtain
the density for the right half of the system,
when we look at the system by setting the origin at the
position of the defect particle.
The density for the left half of the system is obtained by
using the above symmetry as
\begin{equation}
  \label{ri-le}
\langle n_j \rangle_{L,N} (\alpha,\beta)
=
1-\langle n_{L-j+1} \rangle_{L,L-N} (\beta,\alpha),  
\end{equation}
where the dependence of $\langle n_j \rangle_L$ 
on the parameters $\alpha$ and $\beta$ are 
explicitly indicated.

So far, we have been considering the situation 
with a fixed number of particles.
In other words, our discussions have been 
for the canonical ensemble.
We are mainly interested in the physical quantities in the canonical 
ensemble. They can be directly compared with 
the results of the computer simulations because they are usually 
done for fixed particle numbers. 
However, it turns out that computations are much easier
in the grand canonical ensemble.
Since the meaning of chemical potential is 
unclear for our model,
we define the grand canonical ensemble simply 
as a superposition of canonical ensembles.
Let us introduce the fugacity
$\xi^2$ which is associated with the ordinary particles. 
Then the coefficients $P_{\text{GCE}}(\tau_1,\tau_2,\ldots,\tau_L)$
of the normalized stationary state for the grand canonical ensemble
is compactly represented as 
\begin{equation}
  \label{mpa_gce}
  P_{\text{GCE}}(\tau_1,\tau_2,\ldots,\tau_L)
  =
  \frac{1}{Z_L(\xi)}
  \langle W|
  \prod_{j=1}^{L}
  [ \tau_j \xi^2 D 
    +
   (1-\tau_j) E  ] 
  |V\rangle . 
\end{equation}
Here the normalization constant $Z_L(\xi)$ is 
a summation of the coefficients of the unnormalized states.
In matrix notation, this is simply written as
\begin{equation}
  \label{Z_L}
  Z_L(\xi)
  =
  \langle W|C^L|V\rangle ,
\end{equation}
with 
\begin{equation}
  \label{G}
  C
  =
  \xi^2 D +  E .
\end{equation}
We took the fugacity not as $\xi$ but as $\xi^2$.
This is only for the notational simplicity in the 
formulae in the subsequent sections.
But this causes an ambiguity; $\xi$ and $-\xi$ gives
the same fugacity $\xi^2$. In the following we sometimes
give explanation only for $\text{Re}\,\xi>0$.
But it is clear how the $\text{Re}\,\xi<0$ case should be
considered. 
One remark is in order.
If we set $\xi=1$ in (\ref{mpa_gce}), 
$P_{\text{GCE}}(\tau_1,\tau_2,\ldots,\tau_L)$
gives exactly the stationary state of 
the partially ASEP with {\it open  boundary}
condition in which there are particle input 
(resp. output) at the left (resp. right) 
end of the system with rate $\alpha$ (resp. $\beta$)
\cite{DEHP,PASEP}.
In other words, the partially ASEP with open
boundaries can be regarded as a superposition 
of the present one defect particle model with all numbers of
particles $N=0,1,2,\ldots,L$ with equal weight.
This fact allows us to compute the average  
and the variance of the density for the 
partially ASEP with open boundaries.
This is done in the Appendix A.

Before closing the section, we give 
some notations, definitions and 
an example of the representation 
of algebraic relations (\ref{alg}).
First we introduce $q$-shifted factorial,
\begin{subequations}
\begin{align}
  (a;q)_n
  &=
  \label{q-shi-fac-1}
  (1-a)(1-aq)(1-aq^2)\cdots(1-aq^{n-1}),
  \\
  (a;q)_0
  &=
  1.
  \label{q-shi-fac-2}
\end{align}
\end{subequations}
We also define 
\begin{equation}
  \label{q-prod-inf}
  (a;q)_{\infty}
  =
  \prod_{j=0}^{\infty}(1-aq^j),
\end{equation}
for $|q|<1$.
Since products of $q$-shifted factorials appear so often,
we use the notations,
\begin{align}
(a_1,a_2,\cdots,a_k;q)_{\infty}
&=
(a_1;q)_{\infty}(a_2;q)_{\infty}\cdots (a_k;q)_{\infty},
\\
(a_1,a_2,\cdots,a_k;q)_{n}
&=
(a_1;q)_{n}(a_2;q)_{n}\cdots (a_k;q)_{n} .
\end{align}
An example of the representation of the
algebra (\ref{alg}) is given by
\begin{subequations}
\begin{gather}
\label{DE_rep}
D
=
\begin{bmatrix}
1       & \sqrt{1-q} & 0             & 0            & \cdots \\
0       & 1          & \sqrt{1-q^2}  & 0            &     \\
0       & 0          & 1             & \sqrt{1-q^3} &  \\
\vdots  &            &               & \ddots       &\ddots \\
\end{bmatrix},
\quad
E
=
D^T,
\quad
A=|V\rangle \langle W|,
\\
\label{WV_rep}
\langle W| 
= 
\kappa\,_c\langle a|
=
\kappa \left( 1, \frac{a}{\sqrt{(q;q)_1} },
             \frac{a^2}{ \sqrt{(q;q)_2} },\ldots \right),
\quad
|V \rangle 
= 
\kappa \, |b\rangle_c
=
\kappa 
\left(
\begin{matrix}
1\\
\displaystyle\frac{b}{ \sqrt{(q;q)_1}}\\
\displaystyle\frac{b^2}{ \sqrt{(q;q)_2} }\\
\vdots
\end{matrix}
\right),
\end{gather}
\end{subequations}
where the superscript $T$ indicates the transposition.
Notice that the special form of the matrix $A$ ensures
the existence of the trace.
We introduced
\begin{align}
a
&=
-1+\frac{1}{\tilde{\alpha}}
=
-1+\frac{1-q}{\alpha},
\\
b
&=
-1+\frac{1}{\tilde{\beta}}
=
-1+\frac{1-q}{\beta}.
\end{align}
From now on,
we restrict our attention to the case where $a>0$ and $b>0$,
i.e., the case where $\alpha<1-q$ and $\beta<1-q$.
This is mainly for the simplicity of the discussions.
We do not expect there appears another phase transition 
if we consider $a<0$ or $b<0$ case.
The constant $\kappa$ in (\ref{WV_rep}) 
is takes as $\kappa^2=(ab;q)_{\infty}$
so that $\langle W|V\rangle=1$.
An important fact about these matrices is that
they are related to the so-called $q$-orthogonal 
polynomials \cite{AW,AI,GR}. It is this connection that allows us
to compute the physical quantities exactly in the subsequent
sections. The relationship between the matrix product ansatz 
for the partially ASEP  
and the theory of the $q$-orthogonal polynomials was first clarified in 
\cite{PASEP} and has been exploited to study the partially
ASEP with open boundaries in \cite{PASEP2,BECE}.

\Section{The Current}
\label{current}
In this section, we compute the current of the ordinary particles
(\ref{def_J}) and the speed of the defect particle (\ref{def_speed})
in the thermodynamic limit.
We need to know the asymptotic expression of $Z_{L,N}$ when
$L,N\rightarrow \infty$ with $\rho=N/L$ fixed.

\subsection{Exact Formula for $Z_L(\xi)$}
First we give the exact expression of $Z_L(\xi)$ in 
the form of contour integral.
Since the derivation is almost the same as 
that in \cite{PASEP}, the proof is omitted.
The result is 
\begin{equation}
  Z_L(\xi)
  =
  \frac{(q,ab;q)_{\infty}}{4\pi i}
  \int_{C} dz
  \frac{(z^2,z^{-2};q)_{\infty}
        [(1+\xi z)(1+\xi z^{-1})]^L  }
       { (a \xi^{-1} z,a \xi^{-1} z^{-1},
          b \xi      z,b \xi      z^{-1};q)_{\infty}}.
\end{equation}
The contour $C$ of the integral above is such that 
it includes all poles of the type $a \xi^{-1}q^k$ and $b \xi q^k$
with $k=0,1,2,\cdots$  whilest it excludes all poles of the 
type $a^{-1}\xi q^{-k}$ and $b^{-1}\xi^{-1} q^{-k}$
with $k=0,1,2,\cdots$.
When $a \xi^{-1}<1$ and  $b \xi <1$ hold, the contour is simply 
a unit circle and $Z_L(\xi)$ reduces to the single 
integral on the real line. 
But when  $a \xi^{-1}>1$ or $b\xi >1$
there appear other terms as well.
In general, when
\begin{gather}
  \label{}
  a \xi^{-1}  > a \xi^{-1} q > \cdots > a \xi^{-1} q^{n^{(a)}} > 1 > 
  a \xi^{-1}  q^{n^{(a)}+1} > \cdots ,
  \notag\\
  \label{}
  b \xi  > b \xi  q > \cdots > b \xi  q^{n^{(b)}} > 1 > 
  b \xi  q^{n^{(b)}+1} > \cdots ,
\end{gather}
where $n^{(a)}$ and $n^{(b)}$ are some non-negative integers, we have
\begin{equation}
  \label{decomp_ZL}
  Z_L(\xi)=Z_L^{(0)}(\xi)+Z_L^{(a)}(\xi)+Z_L^{(b)}(\xi).
\end{equation}
Explicit expressions for the above three terms read
\begin{align}
  Z_L^{(0)}(\xi)
  &=
  \frac{(q,ab;q)_{\infty}}{2\pi}
  \int_{0}^{\pi} d\theta
  \frac{(e^{2i\theta},e^{-2i\theta};q)_{\infty}
        [(1+\xi e^{i\theta})(1+\xi e^{-i\theta})]^L  }
       {(a \xi^{-1} e^{i\theta},a \xi^{-1} e^{-i\theta},
         b \xi      e^{i\theta},b \xi      e^{-i\theta};q)_{\infty}},
  \\
  \label{}
  Z_L^{(a)}(\xi)
  &=
  (q,ab;q)_{\infty}
  \sum_{j=0}^{n^{(a)}} w_j^{(a)} [(1+a q^j)(1+\xi^2 a^{-1} q^{-j})]^L, 
  \\
  \label{}
  Z_L^{(b)}(\xi)
  &=
  (q,ab;q)_{\infty}
  \sum_{j=0}^{n^{(b)}} w_j^{(b)} [(1+\xi^2 b q^j)(1+b^{-1} q^{-j})]^L, 
\end{align}
with
\begin{align}
  \label{}
  w_j^{(a)}
  &=
  \frac{(a^{-2}\xi^2;q)_{\infty} 
        (a^2\xi^{-2},ab;q)_j
        (1-\xi^{-2}a^2 q^{2j}) 
        \xi^{2j}}
       {(q,ab,a^{-1}b\xi^2;q)_{\infty}
        (q,ab^{-1}\xi^{-2}q;q)_j
        (1-\xi^{-2}a^2) 
        q^{j^2} a^{3j} b^j},
  \\
  \label{}
  w_j^{(b)}
  &=
  \frac{(b^{-2}\xi^{-2};q)_{\infty} 
        (b^2\xi^{2},ab;q)_j
        (1-\xi^{2}b^2 q^{2j}) }
       {(q,ab,a b^{-1}\xi^{-2};q)_{\infty}
        (q,a^{-1} b \xi^{2} q;q)_j
        (1-\xi^{2}b^2) 
        q^{j^2} a^{j} b^{3j} \xi^{2j}}.
\end{align}   
Off course when $a\xi^{-1}<1$ and/or $b\xi<1$, 
we should omit $Z_L^{(a)}$ and/or $Z_L^{(b)}$ 
in (\ref{decomp_ZL}).

The asymptotic behavior of $Z_L^{(0)}(\xi)$ can 
be evaluated by applying the steepest descent method,
whereas the asymptotic behaviors of $Z_L^{(a)}(\xi)$
and $Z_L^{(b)}(\xi)$ are simply given by the $j=0$ terms
in the summation. We find 
\begin{gather}
  \label{ZL0as}
  Z_L^{(0)}(\xi)
  \simeq
  \frac{(q,ab;q)_{\infty} [(1+\xi)(1+\xi^{-1})]^{3/2}}
       {2\sqrt{\pi}(a\xi^{-1},b\xi;q)_{\infty}^2 L^{3/2}}
  (1+\xi)^{2L},
  \\
  \label{ZLAas}
  Z_L^{(a)}(\xi)
  \simeq
  \frac{(\xi^2 a^{-2};q)_{\infty}}
       {(a^{-1}b\xi^2;q)_{\infty}}
  [(1+a)(1+\xi^2a^{-1})]^L ,
  \\
  \label{ZLBas}
  Z_L^{(b)}(\xi)
  \simeq
  \frac{(\xi^{-2} b^{-2};q)_{\infty}}
       {(a b^{-1}\xi^{-2};q)_{\infty}}
  [(1+\xi^2 b)(1+b^{-1})]^L 
\end{gather}
as $L\rightarrow\infty$ when $\text{Re}\,\xi>0$.

\subsection{Asymptotic Formula for $Z_{L,N}$}
In this subsection we study the asymptotic behavior of 
$Z_{L,N}$. 
First we notice 
\begin{equation}
  \label{ZL_ZLN}
  Z_L(\xi)
  = 
  \sum_{N=0}^{L}
  Z_{L,N} \xi^{2N}.
\end{equation}
The average density of particles is given by
\begin{equation}
  \label{den_fu}
  \langle\rho\rangle
  =
  \xi^2
  \lim_{L\rightarrow\infty}
  \frac1L\frac{\partial}{\partial \xi^2}\ln Z_L(\xi).
\end{equation}
Basically we can expect that
each value of the fugacity $\xi^2$
corresponds to each value of the density. 
Then the density-fugacity relation is invertible
and the equivalence of the canonical and the 
grand canonical ensemble holds. For our model,
however, there appears a situation where 
the equivalence of the ensembles fails. 
For this case, we have to go back to (\ref{ZL_ZLN})
and use the inversion of it,
\begin{equation}
  \label{Z_LN}
  Z_{L,N}
  =
  \frac{1}{2\pi i}
  \int_{C_1} d\xi
  \frac{Z_L(\xi)}{\xi^{2N+1}},
\end{equation}
where $C_1$ is a contour which encircles the origin anti-clockwise.
Notice that (\ref{Z_LN}) is always true for a finite $L$
since $Z_L(\xi)$ is a polynomial of degree $2N$. 
This formula also has an advantage that
it allows us to obtain not only the exponent
of the asymptotic behavior of $Z_{L,N}$ 
but also the prefactor of it.

In order to find the relationship between the 
density and the fugacity and derive the 
asymptotic expression of $Z_{L,N}$,
we need to know, for a give value of $\xi$, 
which term in (\ref{decomp_ZL})
gives the main contribution to $Z_L(\xi)$.
For notational convenience we define $\xi_0=\rho/(1-\rho)$.

\vspace{6mm}
\noindent
{\it When $\tilde{\alpha}+\tilde{\beta}>1$ \hspace{2mm}}($ab<1$ ) 
\begin{itemize}
\item Case $0 < \xi < a$

The main contribution to $Z_L(\xi)$ comes from $Z_L^{(a)}$.
The density-fugacity relation (\ref{den_fu}) gives 
$\xi=\sqrt{a \xi_0}\,(\equiv \xi_a)$. In order for this value to be in the region 
$0 < \xi < a$, the density $\rho$ should satisfy
$\rho<1-\tilde{\alpha}<\tilde{\beta}$.
The asymptotic behavior of $Z_{L,N}^{(a)}$ can be obtained by 
applying the steepest descent method.

\item Case $a < \xi < b^{-1}$

The main contribution to $Z_L(\xi)$ comes from $Z_L^{(0)}$.
The density-fugacity relation (\ref{den_fu}) gives 
$\xi=\xi_0$. In order for this value to be in the region 
$a < \xi < b^{-1}$, the density $\rho$ should satisfy
$1-\tilde{\alpha}<\rho<\tilde{\beta}$.
The asymptotic behavior of $Z_{L,N}^{(b)}$ can be obtained by 
applying the steepest descent method.

\item Case $ \xi > b^{-1}$

The main contribution to $Z_L(\xi)$ comes from $Z_L^{(b)}$.
The density-fugacity relation (\ref{den_fu}) gives 
$\xi=\sqrt{\xi_0/b}\,(\equiv \xi_b)$. 
In order for this value to be in the region 
$ \xi > b^{-1}$, the density $\rho$ should satisfy
$1-\tilde{\alpha}<\tilde{\beta}<\rho$.
The asymptotic behavior of $Z_{L,N}^{(0)}$ can be obtained by 
applying the steepest descent method.

\end{itemize}

\vspace{6mm}
\noindent
{\it When $\tilde{\alpha}+\tilde{\beta}<1$ \hspace{2mm}}($ab>1$ ) 
\begin{itemize}
\item  Case $0 < \xi < \sqrt{a/b}$

The main contribution to $Z_L(\xi)$ comes from $Z_L^{(a)}$.
The density-fugacity relation (\ref{den_fu}) gives 
$\xi=\xi_a$. In order for this value to be in the region 
$0 < \xi < \sqrt{a/b}$, the density $\rho$ should satisfy
$\rho<\tilde{\beta}<1-\tilde{\alpha}$.

\item Case $ \xi > \sqrt{a/b}$

The main contribution to $Z_L(\xi)$ comes from $Z_L^{(b)}$.
The density-fugacity relation (\ref{den_fu}) gives 
$\xi=\xi_b$. In order for this value to be in the region 
$ \xi > \sqrt{a/b}$, the density $\rho$ should satisfy
$\tilde{\beta}<1-\tilde{\alpha}<\rho$.

\item Case $\xi = \sqrt{a/b}$

So far we have not seen the values of the fugacity 
which corresponds to the region 
$\tilde{\beta}<\rho<1-\tilde{\alpha}$.
It turns out that this is related to the 
fact that the density-fugacity relation (\ref{den_fu})
breaks down at $\xi = \sqrt{a/b}$.
Notice that $Z_{L,N}^{(a)}(\xi)$ and $Z_{L,N}^{(b)}(\xi)$
in (\ref{ZLAas}) and (\ref{ZLBas}) have a pole at this point. 
Physically, this is related to the existence of a 
shock.
It might seem difficult to know the asymptotic behavior
of $Z_{L,N}$ for this case. But this is not the case.
Let us employ (\ref{Z_LN}) and take 
the contour as a circle $C_{R}$ with the radius $R$.
As long as $L$ is finite, we can change the radius $R$ 
as we wish. To know the asymptotic behavior
of $Z_{L,N}$, we want to use (\ref{ZL0as})-(\ref{ZLBas})
and apply the steepest descent method.
It turns out that the saddle point associated with 
$Z_L^{(a)}(\xi)$ (resp. $Z_L^{(b)}(\xi)$)
 is at $\xi=\xi_a$ (resp. $\xi=\xi_b$) and 
the contribution from $Z_L^{(0)}(\xi)$
is smaller than the others.
Hence, to apply the steepest descent method,
we have to take the radius of the contour as
$R_1=\xi_a ( >\sqrt{a/b})$ for  $Z_L^{(a)}(\xi)$ and 
as $R_2=\xi_b ( <\sqrt{a/b})$ for  $Z_L^{(b)}(\xi)$.
At first let us take $R$ as $R_2$. We have
\begin{align}
  Z_{L,N}
  &\simeq
  \frac{1}{2\pi i}\int_{C_{R_2}}
  \frac{d\xi}{\xi}
  \frac{(a^{-2}\xi^2;q)_{\infty}}
       {(a^{-1}b\xi^2;q)_{\infty}}
  \frac{[(1+a)(1+\xi^2a^{-1})]^L}  
       {\xi^{2N}}
  \notag\\
  &\quad
  +
  \frac{1}{2\pi i}\int_{C_{R_2}}
  \frac{d\xi}{\xi}
  \frac{(b^{-2}\xi^{-2};q)_{\infty}}
       {(ab^{-1}\xi^{-2};q)_{\infty}}
  \frac{[(1+\xi^2 b)(1+b^{-1})]^L}  
       {\xi^{2N}}.    
\end{align}
To apply the steepest descent method to the first term,
we have to modify $C_{R_2}$ to $C_{R_1}$.
Since the integrand of the first term has a pole at $\sqrt{a/b}$, there
appears a contribution from the pole when the contour
is modified. More explicitly, we see
\begin{align}
  \label{}
  &\frac{1}{2\pi i}\int_{C_{R_1}}
  \frac{d\xi}{\xi}
  \frac{(a^{-2}\xi^2;q)_{\infty}}
       {(a^{-1}b\xi^2;q)_{\infty}}
  \frac{[(1+a)(1+\xi^2a^{-1})]^L}  
       {\xi^{2N}}
  \notag\\
  =
  &\frac{1}{2\pi i}\int_{C_{R_2}}
  \frac{d\xi}{\xi}
  \frac{(a^{-2}\xi^2;q)_{\infty}}
       {(a^{-1}b\xi^2;q)_{\infty}}
  \frac{[(1+a)(1+\xi^2a^{-1})]^L}  
       {\xi^{2N}}  
  -
  \frac{(a^{-1}b^{-1};q)_{\infty}(1+a)^L(1+b^{-1})^L}
       {(q;q)_{\infty}(ab^{-1})^N}.
\end{align}
Now we can apply the steepest descent method to the
two integrals.
But it turns out that the contribution from the 
pole gives the asymptotic behavior of $Z_{L,N}$
for this case.

\end{itemize}

Combining the above results, we see that there are
four regions in the $\tilde{\alpha}$-$\tilde{\beta}$ plane, 
in each of which the asymptotic behavior of $Z_{L,N}$ 
in the thermodynamic  limit has a different form. 
We refer to these four phases as Phase 
{\bf A}, {\bf B}, {\bf C} and {\bf D}.
The asymptotic expression of $Z_{L,N}$ in each phase 
is given by 

\begin{itemize}
\item Phase {\bf A}: $\tilde{\alpha}<1-\rho,\,\tilde{\beta}>\rho$
  \begin{equation}
    \label{ZLNA}
    Z_{L,N}
    \simeq
    Z_{L,N}^{(a)}
    \simeq
    \frac{(a^{-1}\xi_0;q)_{\infty}}
         {(b \xi_0;q)_{\infty}}
    \frac{1}{\sqrt{2\pi} \tilde{\alpha}^{L-N} (1-\tilde{\alpha})^N
    \rho^{N+1/2}(1-\rho)^{L-N+1/2}}.      
  \end{equation}

\item Phase {\bf B}: $\tilde{\alpha}>1-\rho,\,\tilde{\beta}<\rho$
  \begin{equation}
    \label{ZLNB}
    Z_{L,N}
    \simeq
    Z_{L,N}^{(b)}
    \simeq
    \frac{(b^{-1}\xi_0^{-1};q)_{\infty}}
         {(a \xi_0^{-1};q)_{\infty}}
    \frac{1}  
         {\sqrt{2\pi} \tilde{\beta}^{N} (1-\tilde{\beta})^{L-N}
      \rho^{N+1/2}(1-\rho)^{L-N+1/2}}.
  \end{equation}  

\item Phase {\bf C}: $\tilde{\alpha}>1-\rho,\,\tilde{\beta}>\rho$
  \begin{equation}
    \label{ZLNC}
    Z_{L,N}
    \simeq
    Z_{L,N}^{(0)}
    \simeq
    \frac{\tilde{\alpha} \tilde{\beta} 
      (\tilde{\alpha}+\tilde{\beta}+1)
      (q,abq;q)_{\infty}}
    {(\tilde{\alpha}+\rho-1)^2 (\tilde{\beta}-\rho)^2
      (\xi_0^{-1}aq,\xi_0 b q;q)_{\infty}^2}
    \frac{1}
    {2\pi L^2 \rho^{2N}(1-\rho)^{2L-2N}}.      
  \end{equation}

\item Phase {\bf D}: $\tilde{\alpha}<1-\rho,\,\tilde{\beta}<\rho$
  \begin{equation}
    \label{ZLND}
    Z_{L,N}
    \simeq
    \frac{(1-\tilde{\alpha}-\tilde{\beta}) (a^{-1}b^{-1}q;q)_{\infty}}
         {(1-\tilde{\alpha})(1-\tilde{\beta})(q;q)_{\infty}}
    \frac{1}{\tilde{\alpha}^{L-N}(1-\tilde{\alpha})^N
             \tilde{\beta}^N(1-\tilde{\beta})^{L-N}}           
    \,\,(= 
    Z_{L,N}^{(d)}).
  \end{equation}

\end{itemize}  

These are simple generalization of the results in \cite{Mallick}
for the totally asymmetric case.

\subsection{$J$ and $v$ in the Thermodynamic Limit}
Now it is straight forward to calculate 
the current of the ordinary particles $J$ and 
the speed of the defect particle $v$ in the thermodynamic limit.
The results are summarized as follows:
\begin{itemize}
\item Phase {\bf A}: 
$J=(1-q)\rho(1-\rho)$ and $v=(1-q)(\tilde{\alpha}-\rho)$.

\item Phase {\bf B}:
$J=(1-q)\rho(1-\rho)$ and $v=(1-q)(1-\tilde{\beta}-\rho)$.

\item Phase {\bf C}:
$J=(1-q)\rho(1-\rho)$ and $v=(1-q)(1-2\rho)$.

\item Phase {\bf D}:
$J= (1-q)[\rho(\tilde{\alpha}-\tilde{\beta})
   +\tilde{\beta}(1-\tilde{\alpha})]$ and 
$v=(1-q)(\tilde{\alpha}-\tilde{\beta})$.

\end{itemize}

\Section{Density Profile}
\label{density}
Now we turn to consider the average density profile.
As in the previous section, we first calculate the 
density in the grand canonical ensemble. 
In the grand canonical ensemble, 
the (unnormalized) average density at site $j$ is defined by 
\begin{align}
  \langle n_j \rangle_{L}^{(u)}(\xi) 
  &=
  \sum_{\tau_1,\tau_2,\ldots,\hat{\tau}_j,\cdots,\tau_L}
  \langle V|
  \prod_{k=1}^{j-1}
  [ \tau_j \xi^2  D 
    +
   (1-\tau_j)E]
  \xi^2 D
  \prod_{k=j+1}^{L}
  [ \tau_j  \xi^2 D 
    +
   (1-\tau_j)E]
  |V\rangle .
\end{align}
To translate the results in the grand canonical ensemble
into those in the canonical ensemble, we need the formula,
\begin{equation}
  \label{den_GCE_CE}
  \langle n_j \rangle_{L,N}^{(u)}
  =
  \frac{1}{2\pi i}\int_{C_1}
  d \xi
  \frac{\langle n_j \rangle_L^{(u)}(\xi)} 
       {\xi^{2N+1}}.  
\end{equation}

Now we explain the main idea how to compute 
$\langle n_j \rangle_L^{(u)}(\xi)$.
Details of the computation are not presented here
since they are similar to those in \cite{PASEP2}.
Since it is easier 
to calculate the density difference than the density itself,
we rewrite the density at site $j$ as
\begin{equation}
  \label{density-decompose}
  \langle n_j \rangle_L
  =
  \sum_{k=j}^{L-1}
  (\langle n_k \rangle_L^{(u)}(\xi) 
   - 
   \langle n_{k+1} \rangle_L^{(u)}(\xi))
  +
  \langle n_L \rangle_L^{(u)}(\xi)  .
\end{equation}
At the  right boundary, we have
\begin{equation}
  \label{d-right}
   \langle n_L \rangle_L^{(u)}(\xi) 
  =
  \frac{\xi^2}{\tilde{\beta}}
  \langle W| C^{L-1}|V\rangle .
\end{equation}
Similar formula can be found for the left boundary as well.
As for the computation of 
$\langle n_k \rangle_L^{(u)}(\xi) 
 - 
 \langle n_{k+1} \rangle_L^{(u)}(\xi)$,
we notice that
\begin{align}
  \langle n_k \rangle_L^{(u)}(\xi)
  -
  \langle n_{k+1} \rangle_L^{(u)}(\xi)
  &=
  \xi^2
  (\langle W| C^{k-1}DC^{L-k}|V\rangle
   -
   \langle W| C^{k}DC^{L-k-1}|V\rangle )
  \notag\\
  &=
  \xi^2
  \langle W|C^{k-1}(DC-CD)C^{L-k-1}|V\rangle
  \notag\\
  &=
  \xi^2
  \langle W|C^{k-1}(DE-ED)C^{L-k-1}|V\rangle.
\end{align}
By using the two facts that $DE-ED$ is a simple 
diagonal matrix and that the resultant series  
can be summed up by using a formula for the
$q$-Hermite polynomials, it turns out to be 
possible to derive an integral expression 
of  $ \langle n_k \rangle_L^{(u)}(\xi)
  -
  \langle n_{k+1} \rangle_L^{(u)}(\xi)$.
If we sum up these terms from $k=j$ to $k=L-1$,
after some computation, we get
\begin{equation}
  \label{I1-I2}
  \sum_{k=j}^{L-1}
  (\langle n_k \rangle_L^{(u)}(\xi) 
   - 
   \langle n_{k+1} \rangle_L^{(u)}(\xi) )  
  =
  I_1+I_2,  
\end{equation}
where
\begin{align}
I_1
&=
\frac{\xi}{4} 
(ab;q)_{\infty}(q;q)_{\infty}^3
\int\frac{d z_1}{2\pi i z_1}
\int\frac{d z_2}{2\pi i z_2}
\notag\\
&\quad
\times
\frac{(z_1^2,z_1^{-2},z_2^2,z_2^{-2};q)_{\infty}
      [(1+\xi z_1)(1+\xi z_1^{-1})]^{L-1}}
     {(a\xi^{-1}z_1,a\xi^{-1} z_1^{-1},qz_1 z_2, q z_1^{-1} z_2^{-1},
       q z_1 z_2^{-1},q z_1^{-1} z_2^{-1}, b\xi z_2,b\xi z_2^{-1};q)_{\infty}
      (z_2 + z_2^{-1} -z_1-z_1^{-1})},
\notag\\
I_2
&=
\frac{\xi}{4} 
(ab;q)_{\infty}(q;q)_{\infty}^3
\int\frac{d z_1}{2\pi i z_1}
\int\frac{d z_2}{2\pi i z_2}
\notag\\
&\quad
\times
\frac{(z_1^2,z_1^{-2},z_2^2,z_2^{-2};q)_{\infty}
      [(1+\xi z_1)(1+\xi z_1^{-1})]^{j-1}
      [(1+\xi z_2)(1+\xi z_2^{-1})]^{L-j} }
     {(a\xi^{-1}z_1,a \xi^{-1}z_1^{-1},qz_1 z_2, q z_1^{-1} z_2^{-1},
       q z_1 z_2^{-1},q z_1^{-1} z_2^{-1}, b\xi z_2,b\xi z_2^{-1};q)_{\infty}
      (z_2 + z_2^{-1} -z_1-z_1^{-1})}.
\end{align}
When $0< a\xi^{-1},b\xi<1$, 
the contours of  $z_1$ and $z_2$ are both unit circles.
For other values of the parameters, 
the contours are modified so that the analyticity of
$I_1$ and $I_2$ are ensured.

Combining the results for $\langle n_j \rangle_L^{(u)}(\xi)$
with the formula (\ref{den_GCE_CE}), 
we obtain the density profile of the ordinary particles 
in the canonical ensemble.  
The density at the right boundary is computed by using (\ref{d-right}).
The integral $I_1$ gives us the density at bulk region,
which turns out to be $\rho$ except for the phase {\bf D}.
Off course this is consistent with the fact 
that we are now dealing with the system with the fixed density $\rho$.
The integral $I_2$ contains the information about 
how the density decays near the right boundary.
The density near the left boundary can be found by 
noticing the symmetry (\ref{ri-le}).
The results are summarized in the following.

\renewcommand{\labelenumi}{5.\Roman{enumi}}
\vspace{5mm}
\noindent
{\bf Phase A} 
  ($\tilde{\alpha}<1-\rho$ and $\tilde{\beta}>\rho$;
  $a>\xi_0$ and $b<1/\xi_0$)      
 
  \noindent
  The average density in the bulk region takes the constant 
  value $\rho$.
  The densities at boundaries are given by
  \begin{equation}
    \label{}
    \langle n_1 \rangle_L
    = 
    \rho,
    \quad
    \langle n_L \rangle_L
    = 
    \frac{1-\tilde{\alpha}}{\tilde{\beta}}\rho.    
  \end{equation}
  Notice that the density near the left boundary does not change and takes the 
  constant value $\rho$.
  Computation of the density near the right boundary shows that 
  the phase {\bf A} subdivides into three phases.

  \begin{itemize}
  \item Phase {\bf A}$_1$
    
    This phase corresponds to 
    \begin{equation}
      \rho < \tilde{\beta} 
      < \frac{\rho}{\rho+(1-\rho)q},
      \quad 
      \tilde{\beta} < 
      \frac{\sqrt{(1-\tilde{\alpha})\rho}}
           {\sqrt{\tilde{\alpha}(1-\rho)}+\sqrt{(1-\tilde{\alpha})\rho}}
    \end{equation}
    in the $\tilde{\alpha}$-$\tilde{\beta}$ plane and to 
    \begin{equation}
      q/\xi_0<b<1/\xi_0, 
      \quad
      b>1/\sqrt{a\xi_0}
    \end{equation}
    in the $a$-$b$ plane. 
    Let us denote the distance from the right boundary as 
    $l=L-j+1$ from now on.
    Then the density near the right boundary decays exponentially as
    \begin{equation}
      \label{density-A1}
      \langle n_j \rangle_L
      =
      \rho
      -
      \frac{(ab^2\xi_0-1)(q,a^{-1}b^{-2}\xi_0^{-1}q;q)_{\infty}}
           {(1+ab\xi_0)(1+b)(a^{-1}b^{-1}q,b^{-1}\xi_0^{-1}q;q)_{\infty}}
      \exp[-l/r]
    \end{equation}
    with the correlation length,
    \begin{equation}
      r^{-1}
      =
      -\ln 
      \left[
      \frac{\tilde{\alpha}(1-\rho)}{1-\tilde{\beta}}
            +
      \frac{(1-\tilde{\alpha})\rho}{\tilde{\beta}}
      \right].
    \end{equation}

  \item Phase {\bf A}$_2$

    This phase corresponds to 
    \begin{equation}
      \frac{(1-\rho)q^2}{\rho+(1-\rho)q^2} 
      < 
      \tilde{\alpha} 
      < 
      1-\rho,
      \quad 
      \tilde{\beta} > 
      \frac{\sqrt{(1-\tilde{\alpha})\rho}}
           {\sqrt{\tilde{\alpha}(1-\rho)}+\sqrt{(1-\tilde{\alpha})\rho}}
    \end{equation}
    in the $\tilde{\alpha}$-$\tilde{\beta}$ plane and to 
    \begin{equation}
      \xi_0<a<\xi_0 q^{-2}, 
      \quad
      b < 1/\sqrt{a\xi_0}
    \end{equation}
    in the $a$-$b$ plane. 
    The density profile near the right boundary decays
    exponentially as
    \begin{equation}
      \label{density-A2}
      \langle n_j \rangle_L
      =
      \rho
      -
      \frac{\sqrt{\xi_0}(1+a\xi_0)(ab,b\xi_0;q)_{\infty} (q;q)_{\infty}^4}
           {2\sqrt{a\pi}(\sqrt{a\xi_0^{-1}}q,\sqrt{a^{-1}\xi_0},
            b\sqrt{a\xi_0};q)_{\infty}^2}
      \frac{\exp[-l/r]}
           {l^{\frac32}},
    \end{equation}
    with  
    the correlation length, 
    \begin{equation}
      r^{-1}
      =
      -2\ln [\sqrt{\tilde{\alpha}(1-\rho)}+\sqrt{(1-\tilde{\alpha})\rho}].
    \end{equation}
    The decay of the density is not purely exponential 
    but with algebraic corrections.

  \item Phase {\bf A}$_3$

    This phase corresponds to 
    \begin{equation}
      \tilde{\alpha} <
      \frac{(1-\rho)q^2}{\rho+(1-\rho)q^2}, 
      \quad 
      \tilde{\beta} > 
      \frac{\rho}
           {\rho+(1-\rho)q}
    \end{equation}
    in the $\tilde{\alpha}$-$\tilde{\beta}$ plane and to 
    \begin{equation}
      a > \xi_0 q^{-2}, 
      \quad
      b < q \xi_0^{-1}
    \end{equation}
    in the $a$-$b$ plane. 
    The density profile near the right boundary decays
    exponentially as       
    \begin{equation}
      \label{density-A3}
      \langle n_j \rangle_L
      =
      \rho
      -
      \frac{\xi_0(1-ab)(1-a^{-1}\xi_0q^{-2})}
           {(1-b\xi_0q^{-1})(1+aq)(1+\xi_0q^{-1})}
      \exp[-l/r]
    \end{equation}
    with the correlation length, 
    \begin{equation}
      r^{-1}
      =
      \ln \frac{q}
               {[\tilde{\alpha}+(1-\tilde{\alpha})q]
                [\rho + (1-\rho)q]}. 
    \end{equation}
  \end{itemize}       

\noindent
{\bf Phase B}
  ($\tilde{\alpha}>1-\rho$ and $\tilde{\beta}<\rho$;
  $a<\xi_0$ and $b>1/\xi_0$)      

  \noindent 
  This phase is symmetric to the Phase {\bf A}.
  All results can be obtained by using the symmetry 
  (\ref{symmetry}). 
  The average density in the bulk region takes the constant 
  value $\rho$ and does not change near the right boundary.
  The densities at boundaries are given by
  \begin{equation}
    \label{}
    \langle n_1 \rangle_L
    = 
    1-\frac{1-\tilde{\beta}}{\tilde{\alpha}}(1-\rho),
    \quad
    \langle n_L \rangle_L
    = 
    \rho.    
  \end{equation} 

  \begin{itemize}
  \item Phase {\bf B}$_1$
    
    This phase corresponds to 
    \begin{equation}
      1-\rho < \tilde{\alpha} 
      < \frac{1-\rho}{1-\rho+\rho q},
      \quad 
      \tilde{\alpha} < 
      \frac{\sqrt{(1-\tilde{\beta})(1-\rho)}}
           {\sqrt{\tilde{\beta}\rho}+\sqrt{(1-\tilde{\beta})(1-\rho)}}
    \end{equation}
    in the $\tilde{\alpha}$-$\tilde{\beta}$ plane and to 
    \begin{equation}
      \xi_0 q < a < \xi_0, 
      \quad
      b > \frac{\xi_0}{a^2}
    \end{equation}
    in the $a$-$b$ plane. 
    The density near the left boundary decays exponentially 
    with the correlation length,
    \begin{equation}
      r^{-1}
      =
      -\ln 
      \left[
      \frac{\tilde{\beta}\rho}{1-\tilde{\alpha}}
            +
      \frac{(1-\tilde{\beta})(1-\rho)}{\tilde{\alpha}}
      \right].
    \end{equation}
    
  \item Phase {\bf B}$_2$

    This phase corresponds to 
    \begin{equation}
      \frac{\rho q^2}{1-\rho+\rho q^2} 
      < 
      \tilde{\alpha} 
      < 
      1-\rho,
      \quad 
      \tilde{\alpha} > 
      \frac{\sqrt{(1-\tilde{\beta})(1-\rho)}}
           {\sqrt{\tilde{\beta}\rho}+\sqrt{(1-\tilde{\beta})(1-\rho)}}
    \end{equation}
    in the $\tilde{\alpha}$-$\tilde{\beta}$ plane and to 
    \begin{equation}
      \xi_0^{-1}<b<\xi_0^{-1} q^{-2}, 
      \quad
      a < b<\frac{\xi_0}{a^2}
    \end{equation}
    in the $a$-$b$ plane. 
    The density profile near the left boundary decays
    exponentially 
    with the correlation length, 
    \begin{equation}
      r^{-1}
      =
      -2\ln 
      \left[
        \sqrt{\tilde{\beta}\rho}+\sqrt{(1-\tilde{\beta})(1-\rho)}
      \right].
    \end{equation}

  \item Phase {\bf B}$_3$

    This phase corresponds to 
    \begin{equation}
      \tilde{\beta} <
      \frac{\rho q^2}{1-\rho+\rho q^2}, 
      \quad 
      \tilde{\alpha} > 
      \frac{1-\rho}
           {1-\rho+\rho q}
    \end{equation}
    in the $\tilde{\alpha}$-$\tilde{\beta}$ plane and to 
    \begin{equation}
      b > \xi_0^{-1} q^{-2}, 
      \quad
      a < \xi_0 q
    \end{equation}
    in the $a$-$b$ plane. 
    The density profile near the left boundary decays
    exponentially 
    with the correlation length, 
    \begin{equation}
      r^{-1}
      =
      \ln \frac{q}
               {[\tilde{\beta}+(1-\tilde{\beta})q]
                [1-\rho + \rho q]}. 
    \end{equation}
  \end{itemize}

\noindent       
{\bf Phase {\bf C}}
  ($1-\tilde{\alpha} < \rho< \tilde{\beta}$; 
   $a<\xi_0$ and $b<\xi_0^{-1}$)
  
  \noindent
  The average density in the bulk region takes the constant   
  value $\rho$.
  The densities at boundaries are given by
  \begin{equation}
    \label{}
    \langle n_1 \rangle_L
    = 
    1-\frac{1}{\tilde{\alpha}}(1-\rho)^2,
    \quad
    \langle n_L \rangle_L
    = 
    \frac{\rho^2}{\tilde{\beta}}.    
  \end{equation} 
  The average density decays near the right boundary as
  \begin{equation}
    \label{density-max}
    \langle n_j \rangle_L
    =
    \rho 
    -
    \sqrt{\frac{\rho(1-\rho)}{\pi}}
    \frac{1}{l^{\frac12}}.
  \end{equation}
  The density decays algebraically and hence the correlation length is 
  infinite.
  The density decay at the left boundary can be obtained from 
  the symmetry relation (\ref{ri-le}).
  This is the same as the result for the $q=0$ case in \cite{Mallick}.

\vspace{4mm}  
\noindent  
{\bf Phase {\bf D}}
  ($\tilde{\alpha} <1- \rho$ and $\tilde{\beta}<\rho$;
   $a>\xi_0$ and $b>\xi_0^{-1}$)
  
  \noindent
  In this phase, there is a shock. Hence the average density
  at bulk region is not a constant. 
  For the densities at the boundaries, we have
  \begin{equation}
    \label{}
    \langle n_1 \rangle_L
    = 
    \tilde{\beta},
    \quad
    \langle n_L \rangle_L
    = 
    1-\tilde{\alpha}.    
  \end{equation} 
  There are two regions in the system. One is the 
  low density region with the density $\tilde{\beta}$
  at left side and the other is the high density 
  region with the density $1-\tilde{\alpha}$ at 
  right side. The two regions are separated by 
  a sharp interface, which we call a shock.
  When $L$ is large, the average density profile
  is described by a continuous function $n(x)$ 
  in terms of the rescaled variable $x=j/L$ 
  ($0\leq x\leq 1$).
  The derivative of $n(x)$ is found to be
  \begin{equation}
    \label{dndx}
    \frac{d n(x)}{d x}
    \simeq
    C \, \exp[L\cdot F(x)],
  \end{equation}
  where 
  \begin{subequations}
  \begin{align}
    \label{def_F}
    F(x)
    &=
    x\ln[(1+a)(1+a^{-1}G(x))]
    +
    (1-x)\ln[(1+b\, G(x))(1+b^{-1})],
    \\
    \label{def_G}
    G(x)
    &=
    \frac{-x+\rho-ab(1-x-\rho) + 
          \sqrt{ [(x-\rho)^2+ab(1-x-\rho)]^2 + 4ab\rho(1-\rho) }}
         {2 b (1-\rho)}.
  \end{align}
  \end{subequations}
  The constant $C$ in (\ref{dndx}) is determined by the 
  normalization condition,
  \begin{equation}
    \label{n_norm}
    \int_0^1 \frac{d n(x)}{d x} dx
    =
    1-\tilde{\alpha}-\tilde{\beta}.
  \end{equation}
  It turns out that the function $F(x)$ takes
  the maximum value at 
  $x_0=(1-\tilde{\alpha}-\rho)/(1-\tilde{\alpha}-\tilde{\beta})$.
  If we expand $F(x)$ around $x=x_0$ to the second order 
  and approximate (\ref{dndx}) by a Gaussian, we find
  \begin{equation}
    \label{dndx_ap}
    \frac{d n(x)}{dx}
    \simeq
    \sqrt{\frac{L}{2\pi[ \rho(\tilde{\alpha}-\tilde{\beta})
                        +\tilde{\beta}(1-\tilde{\alpha})]}}
    (1-\tilde{\alpha}-\tilde{\beta})^2
    \exp \left(
            -\frac{L (1-\tilde{\alpha}-\tilde{\beta})^2 (x-x_0)^2}
                  {2 [ \rho(\tilde{\alpha}-\tilde{\beta})
                        +\tilde{\beta}(1-\tilde{\alpha})]}
         \right).
  \end{equation}
  This is essentially the same expression as that for the totally
  asymmetric case \cite{Mallick}.
  But it should be remarked that (\ref{dndx_ap}) 
  is only a consequence of the above approximation;
  the correct formula is given by (\ref{dndx}) with
  (\ref{def_F}) and (\ref{def_G}).

\vspace{8mm}
We thus obtain the phase diagram shown in Fig. 1.
As we have seen, the phase {\bf A} (resp. {\bf B}) for 
the current subdivides into three phases 
{\bf A}$_1$, {\bf A}$_2$ and {\bf A}$_3$
(resp. {\bf B}$_1$, {\bf B}$_2$ and {\bf B}$_3$).
This phase diagram has a richer structure than 
that for the totally asymmetric case \cite{Mallick}
for which the phase {\bf A} (resp. {\bf B}) for 
the current subdivides into only two phases 
and the phases {\bf A}$_3$ and {\bf B}$_3$
are not observed. 
The situation is analogous to that for the
ASEP with open boundaries \cite{PASEP,PASEP2}.

\Section{Concluding Remarks}
\label{conc}
In this article, 
we have considered the partially asymmetric simple
exclusion process on a ring with one defect particle.
We have computed the current of the ordinary particles,
the speed of the defect particle and the density profile
of the ordinary particles seen from the defect particle.
The main result of this article is the identification
of the phase diagram for the correlation length 
shown in Fig. 1.
It turns out that the phase diagram has a richer
structure than that for the totally asymmetric case.

There are several possible applications and 
generalizations of the analysis in this article.
First there are several models for which the partially
asymmetric case has not been solved.
For instance we can generalize the analysis in \cite{DJLS}  
for the totally asymmetric case to the partially
asymmetric case.
Second, we didn't study the case where $q\geq 1$ 
in this article since our main interest was the 
phase structure of the model and it is obvious that 
there is a phase transition at $q=1$. 
But it would be interesting to apply the similar analysis
to this case and compute the physical quantities exactly.
In addition, if we consider the limit $q\rightarrow 1$
carefully, the crossover behavior should be observed
since this limit is associated with the change of the universality of the
model from KPZ universality class to EW universality class
\cite{Kim}. 
Here we only remark that the corresponding ASEP with open boundaries have been
considered in \cite{SEP} for $q=1$ case and 
in \cite{BECE} for $q>1$ case.
Lastly it would be also interesting to generalize our analysis
to the multi-species models
\cite{EFGM,EKKM,AHR98-1,ADR,MMR}.
Compared to the ASEP,
much less is known about these models.
Several investigations are now in progress \cite{RSS}.
The results about these will be reported elsewhere.

\section*{Acknowledgment}
The author would like to thank P. Deift, E. R. Speer 
for fruitful discussions and comments. 
He also thanks the continuous encouragement of 
M. Wadati.
The author is a Research Fellow of the Japan Society
for the Promotion of Science.

\appendix
\renewcommand{\thesection}{Appendix \Alph{section}}
\renewcommand{\theequation}{\Alph{section}.\arabic{equation}}

\section{Density Fluctuation for the Partially \\ ASEP with Open
Boundaries}
In the main text of this article, we are mainly interested in the system
with a fixed number of particles. 
The physical quantities are computed for a given value of
the density $\rho$.
When we consider the ASEP with open boundaries, however,
the situation becomes different.
The system is connected to particle reservoirs at boundaries;
the total number of particles can fluctuate.
As pointed out in the main text, the stationary state of 
the partially ASEP with open boundaries is obtained by 
setting $\xi=1$ in (\ref{mpa_gce}).
Moreover the probability 
that the partially ASEP with open boundaries has $N$ particles 
is given by $Z_{L,N}/Z_L(\xi=1)$.
By using the asymptotic expressions of 
$Z_L(\xi=1)$ in \cite{PASEP} and those of 
$Z_{L,N}^{(f)}$ with $f=0,a,b,d$ in (\ref{ZLNA})-(\ref{ZLND}),
it is possible to obtain the asymptotic behavior of 
the probability measure of the particle density 
when $L\rightarrow\infty$.
In particular, we can compute the average 
and the variance of the density for the
partially ASEP with open boundaries.
In this appendix, we do not explain even the basic properties of the 
model and use some terminologies without definition. 
See \cite{DDM,DE,SD,DEHP,PASEP,PASEP2} for instance.

First we study how the asymptotic expression of $Z_{L,N}$
in (\ref{ZLNA})-(\ref{ZLND}) looks like as a function of
$0<\rho<1$. 
As an example we consider $Z_{L,N}^{(a)}$.
Let us rewrite (\ref{ZLNA}) as
\begin{equation}
  \label{ZLNA_S}
  Z_{L,N}^{(a)}
  \simeq
  \frac{(a^{-1}\xi_0;q)_{\infty}}
  {\sqrt{2\pi\rho(1-\rho)}(b \xi_0;q)_{\infty}}
  \exp[L\cdot S(\rho)],
\end{equation}
with
\begin{equation}
  \label{}
  S(\rho)
  =
  -\ln
  \left[
    \tilde{\alpha}^{1-\rho}(1-\tilde{\alpha})^{\rho}
    \rho^{\rho}(1-\rho)^{1-\rho}
  \right].
\end{equation}
Then it is easy to see
\begin{equation}
  \label{}
  \frac{\partial S}{\partial \rho}
  =
  \ln
  \left[
    \frac{\tilde{\alpha}(1-\rho)}
         {(1-\tilde{\alpha})\rho}
  \right]
\end{equation}
becomes zero at $\rho=\tilde{\alpha}$. In addition, we have
\begin{equation}
  \label{}
  \frac{\partial^2 S}{\partial \rho^2}
  =
  -\frac{1}{\rho(1-\rho)} \quad (<0)
\end{equation}
for $0<\rho<1$. 
Hence $Z_{L,N}^{(a)}$ takes a single maximum
value at $N=\tilde{\alpha} L$. 
If we expand $S(\rho)$ around $\rho=\tilde{\alpha}$ and
substitute it into (\ref{ZLNA_S}), we find
\begin{equation}
  \label{ZLNA_G}
  Z_{L,N}^{(A)}
  \simeq
  \left[
    \frac{1}{\tilde{\alpha}(1-\tilde{\alpha})}
  \right]^L
  \exp 
  \left[
    -\frac{L (\rho-\tilde{\alpha})^2}
          {2\tilde{\alpha}(1-\tilde{\alpha})} 
  \right] ,
\end{equation}
when $\rho\sim\tilde{\alpha}$.
The width of the peak if of order $L^{-1/2}$.
As $L$ goes to infinity,  we expect that
the probability measure of the density is 
given by a Dirac measure $\delta_{\tilde{\alpha}}$.
Similarly $Z_{L,N}^{(b)}$ takes a maximum value
at $\rho=1-\tilde{\beta}$; $Z_{L,N}^{(0)}$
takes a maximum value at $\rho=1/2$.
Now we explain the asymptotic behavior of the
probability measure of the density and compute
the average and the variance of the density.

\begin{itemize}
\item low density phase 
  ($\tilde{\alpha}<\frac12,\,\tilde{\beta}>\tilde{\alpha}$)

  From the results in the main text, we see the following.
  For small densities which satisfy $\rho<1-\tilde{\alpha}$
  and $\rho<\tilde{\beta}$, the asymptotic behavior of $Z_{L,N}$
  is determined by $Z_{L,N}^{(a)}$. For larger values of $\rho$, 
  we have $Z_{L,N}\simeq Z_{L,N}^{(0)}$ when 
  $1-\tilde{\alpha}<\rho<\tilde{\beta}$ and
  $Z_{L,N}\simeq Z_{L,N}^{(b)}$ when 
  $\rho>1-\tilde{\alpha}$. As $L$ goes to infinity, however, 
  the main contribution comes from $Z_{L,N}^{(a)}$ in this phase.

  By using (\ref{ZLNA_G}), 
  we compute the average and the variance of the density to get 
  \begin{gather}
    \langle \rho \rangle 
    =
    \tilde{\alpha},
    \\
    \langle (\rho-\langle \rho \rangle)^2 \rangle
    \simeq
    \frac{\tilde{\alpha}(1-\tilde{\alpha})}{L}.
  \end{gather}
  The variance of the density goes to zero as $L\rightarrow\infty$.

\item high density phase 
  ($\tilde{\beta}<\frac12,\,\tilde{\alpha}>\tilde{\beta}$)  
   
  For large densities which satisfy $\rho >1- \tilde{\alpha}$
  and $\rho > \tilde{\beta}$, the asymptotic behavior of $Z_{L,N}$
  is determined by $Z_{L,N}^{(a)}$. For smaller values of $\rho$, 
  we have $Z_{L,N}\simeq Z_{L,N}^{(0)}$ when 
  $1-\tilde{\alpha}<\rho<\tilde{\beta}$ and
  $Z_{L,N}\simeq Z_{L,N}^{(a)}$ when 
  $\rho<1-\tilde{\alpha}$ and $\rho < \tilde{\beta}$. 
  As $L$ goes to infinity, however, 
  the main contribution comes from $Z_{L,N}^{(b)}$ in this phase.

  The average and the variance  of the density are found to be
  \begin{gather}
    \langle \rho \rangle 
    =
    1-\tilde{\beta},
    \\
    \langle (\rho-\langle \rho \rangle)^2 \rangle
    \simeq
    \frac{\tilde{\beta}(1-\tilde{\beta})}{L}.
  \end{gather}

\item maximal current phase 
  ($\tilde{\alpha}>\frac12,\,\tilde{\beta}>\frac12$)  
   
  For small densities which satisfy $\rho >1- \tilde{\alpha}$, 
  we have $Z_{L,N}\simeq Z_{L,N}^{(a)}$;
  For large densities which satisfy $\rho >\tilde{\beta}$, 
  we have $Z_{L,N}\simeq Z_{L,N}^{(b)}$.
  In the middle region of densities which satisfy
  $1-\tilde{\alpha} < \rho < \tilde{\beta}$, 
  the asymptotic behavior of $Z_{L,N}$
  is determined by $Z_{L,N}^{(0)}$. 
  As $L$ goes to infinity,  
  the main contribution comes from $Z_{L,N}^{(0)}$ in this phase.

  The average and the variance of the density are found to be
  \begin{gather}
    \langle \rho \rangle 
    =
    \frac12,
    \\
    \langle (\rho-\langle \rho \rangle)^2 \rangle
    \simeq
    \frac{1}{8L}.
  \end{gather}
  This formula agrees with the one derived in \cite{DE}
  for the special case $q=0,\alpha=\beta=1$.

\item coexistence line 
  ($\tilde{\alpha}=\tilde{\beta} < \frac12$)  
   
  For small densities which satisfy $\rho < \tilde{\alpha}$, 
  we have $Z_{L,N}\simeq Z_{L,N}^{(a)}$;
  For large densities which satisfy $\rho >1-\tilde{\alpha}$, 
  we have $Z_{L,N}\simeq Z_{L,N}^{(b)}$.
  In the middle region of densities which satisfy
  $\tilde{\alpha} < \rho < 1-\tilde{\alpha}$, 
  the asymptotic behavior of $Z_{L,N}$
  is determined by $Z_{L,N}^{(d)}$. 
  When we set $\tilde{\alpha}=\tilde{\beta}$ in 
  (\ref{ZLND}), we notice that $Z_{L,N}^{(d)}$ does
  not depend on $\rho$.  
  This implies that the probability measure of the density
  does not have a peak. 
  In fact the probability measure $w(\rho)d\rho$ in the 
  thermodynamic limit is 
  \begin{equation}
    \label{}
    w(\rho)
    =
    \begin{cases}
      \frac{1}{1-2\tilde{\alpha}} 
         &  \tilde{\alpha} < \rho < 1-\tilde{\alpha} \\
      0 & \text{otherwise}  
    \end{cases}.
  \end{equation}
  Hence we obtain 
  \begin{gather}
    \langle \rho \rangle 
    =
    \frac12,
    \\
    \langle (\rho-\langle \rho \rangle)^2 \rangle
    \simeq
    \frac{(1-2 \tilde{\alpha})^2}{12}.
  \end{gather}

  Notice that the variance of the density is finite for this case.
  This is related to the fact that we see a shock on the 
  coexistence line. The position of the shock can be anywhere
  \cite{SD,KSKS,DGLS}.
  This leads to a macroscopic change of the total number of
  particles or the finite value  of the variance of the
  density. 
  
\end{itemize}

\newpage
\begin{large}
\noindent
Figure Captions
\end{large}

\vspace{10mm}
\noindent
Fig. 1 : 
Phase diagram for the current and the correlation length.
The phase boundaries for the current are represented by
the thick solid lines; those for the correlation length
are represented by the thin solid lines.
The dotted lines are not the phase boundaries.
They are drawn for  convenience.

\newpage
\begin{picture}(500,500)

\put(60,100){\includegraphics{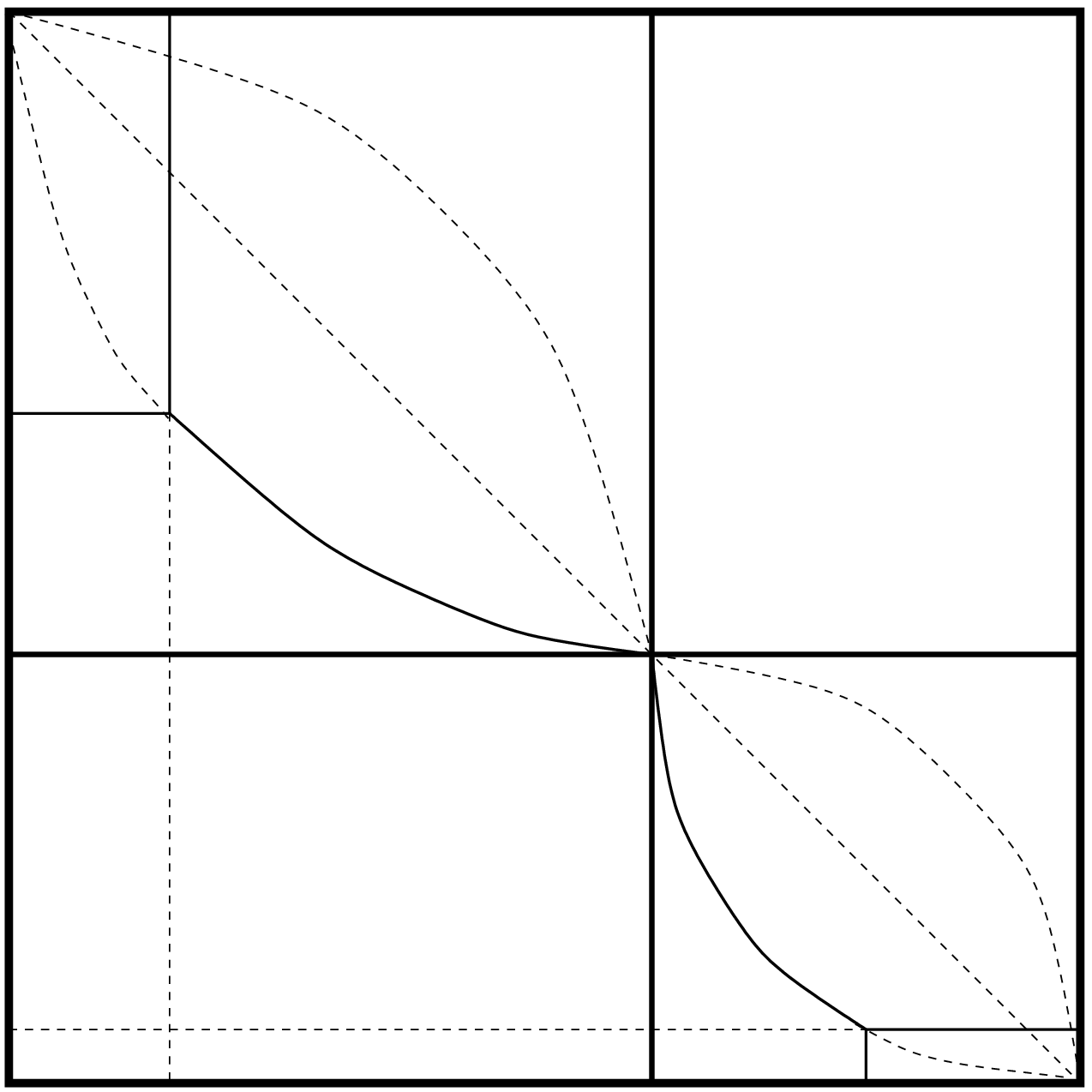}}

\put(420,60){\scalebox{1.5}{$\tilde{\alpha}$}}
\put(17,450){\scalebox{1.5}{$\tilde{\beta}$}}

\put(50,90){$0$}

\put(98,80){$\frac{(1-\rho)q^2}{\rho+(1-\rho)q^2}$}
\put(267,80){$1-\rho$}
\put(337,80){$\frac{1-\rho}{1-\rho+\rho q}$}
\put(425,80){$1$}

\put(10,120){$\frac{\rho q^2}{1-\rho+\rho q^2}$}
\put(35,245){$\rho$}
\put(10,327){$\frac{\rho}{\rho+(1-\rho) q}$}
\put(40,460){$1$}

\put(120,270){\scalebox{1.5}{{\bf A}$_1$}}
\put(180,350){\scalebox{1.5}{{\bf A}$_2$}}
\put(90,380){\scalebox{1.5}{{\bf A}$_3$}}

\put(293,130){\scalebox{1.5}{{\bf B}$_1$}}
\put(360,180){\scalebox{1.5}{{\bf B}$_2$}}
\put(355,135){\scalebox{1.5}{{\bf B}$_3$}}
\put(370,130){\vector(1,-1){15}}

\put(347,350){\scalebox{1.5}{{\bf C}}}
\put(170,170){\scalebox{1.5}{{\bf D}}}

\end{picture}

\end{document}